# Overcoming the curse of dimensionality: Enabling multi-layer photon transport with recurrent neural network

Daniel Carne[a], Ziqi Guo[a], Xiulin Ruan[a,*]

[a] School of Mechanical Engineering, Purdue University, West Lafayette, IN 47907, USA

[*] Corresponding Author: ruan@purdue.edu

**Abstract**

Monte Carlo simulations are commonly used to calculate photon reflectance, absorptance, and transmittance of multi-layer scattering and absorbing media, but they can quickly become prohibitively expensive as the number of layers increases. In this study, we show that although a plain neural network suffers from the curse of dimensionality and fails to yield acceptable predictions of multilayer media, we introduce a recurrent neural network (RNN) trained on the same Monte Carlo simulation dataset to achieve accurate prediction with great acceleration. Our RNN architecture solves the curse of dimensionality by keeping the number of inputs into the network constant for any number of layers. We demonstrate the general applicability with three diverse case studies of multilayer architectures: tissue, radiative cooling paint, and atmospheric clouds, achieving 1-2 orders of magnitude acceleration over Monte Carlo simulations while providing up to one order of magnitude less error than a plain neural network. This recurrent neural network approach enables affordable photon multi-layer modeling, optimization, and high throughput screening for broad applications across dosimetry, atmospheric studies, and spectrally selective radiative coatings.

**Introduction**

Radiative transport simulations through scattering and absorbing media are utilized in many fields, such as neutron transport in nuclear engineering [1], photon transport in heat transfer [2] and climate research [3], and imaging and dosimetry in biomedical applications [4]. These radiative transfer simulations commonly use either the Monte Carlo method or finite volume/difference methods. While often less efficient, the Monte Carlo method is a popular choice because the algorithm for solving is straightforward and highly parallelizable, the prediction is unbiased, and the uncertainty can be easily related to the number of particles modeled [5]. Example applications of Monte Carlo simulations include modeling light propagation in tissue [6], optimizing particle sizes in radiative cooling paints [7], [8], [9], and modeling light scattering and absorption through clouds in the atmosphere [10], [11]. To solve these problems, several Monte Carlo open-source codes have been published and widely used for multi-layer plane-parallel radiative transport, such as MCML by Wang et al. [12]. MCML models radiative transport through multi-layer planar media, which are commonly seen in modeling tissues, the atmosphere, and thin nanocomposite coatings. However, Monte Carlo simulations for multi-layer media can be very computationally intensive, especially for highly scattering media, which can limit optimization [8], high throughput screening [13], and inverse parameter estimation [14].

Recently, regression machine learning techniques have been used as surrogate models to significantly accelerate simulations across many fields including Computational Fluid Dynamics (CFD) [15], [16], phonon transport [17], as well as Monte Carlo radiative transport [18], [19]. Peng et al. used a Convolutional Neural Network (CNN) to denoise a dose map created with Monte Carlo simulations, achieving a 76-fold speedup over high resolution Monte Carlo simulations [20]. Hokr and Bixler used a neural network to solve the inverse Monte Carlo problem, predicting the optical properties of a single layer medium from the spectral response [21]. Neural networks have also directly replaced Monte Carlo simulations for dose maps by predicting where radiation is absorbed in tissue [22], [23]. Our previous work predicted the spectral response (reflectance, absorptance, and transmittance) of single layer media with a fully connected neural network achieving significant speedups of 1-3 orders of magnitude while making it applicable to a broad range of scattering and absorbing materials [24].

However, the fully connected neural network method previously used cannot achieve the same level accuracy for multi-layer spectral response predictions due to the curse of dimensionality. The previous network takes five inputs, the refractive index ($n$), absorption coefficient ($\mu_a$), scattering coefficient ($\mu_s$), asymmetry parameter ($g$), and layer thickness ($t$), but can be non-dimensionalized to reduce the number of inputs to four. Trained on a dataset of 41,000 Monte Carlo simulations, this gives a data density of about 14 datapoints per input dimension ($14^4 \approx 41,000$). To use this same method to train a neural network on three-layer media, for example, would require 12 inputs (4 inputs per layer). If the same size dataset of 41,000 Monte Carlo simulations is used, this would give a data density of only about 2.4 datapoints per input dimension ($2.4^{12} \approx 41,000$). This is too sparse to retain the accuracy required for many applications. If the original data density is to be maintained, a dataset of $14^{12} \approx 5.7 * 10^{13}$ simulations is needed, which is not feasible. This is known as the curse of dimensionality, or the Hughes phenomenon [25], which states that as the number of inputs grows, the amount of data needed grows exponentially. Methods to deal with the curse of dimensionality generally focus on feature selection and decreasing the number of inputs required. This is commonly done through removing highly correlated inputs by analyzing Pearson or Spearman correlation coefficients [26], [27] or through Principal Component Analysis (PCA) [28]. These methods will not benefit radiative transfer because the material properties cannot be further reduced, and each layer's properties are independent of one another.

To address this challenge, in this study we demonstrate a Recurrent Neural Network (RNN) approach for accelerated prediction of the spectral response, including reflectance, absorptance, and transmittance, in multi-layer media. We show the RNN architecture provides three key benefits over the plain NN previously used. First, by recursively inputting a layer's optical properties, the total number of input dimensions remains the same as the single layer case instead of increasing with the number of layers as a plain NN would. This solves the curse of dimensionality, and as we show later, allows the RNN to perform significantly better than a plain NN on the same size dataset. Second, the architecture allows for prediction of multilayer structures containing any number of layers, even if the number of layers is outside the maximum number of layers in the training dataset. Third, the RNN can train on the entire dataset comprised of media with varying numbers of layers, providing improved accuracy. The plain NN cannot achieve this as separate NNs with different architectures are required to train for media with differing numbers of layers due to the varying number of inputs. This RNN architecture provides significant speedups of up to 2 orders of magnitude over Monte Carlo simulations, allowing for accelerated optimization, high throughput screening, and spectral response predictions for many applications. Additionally, the trained RNN parameters and inferencing code is open sourced on GitHub (https://github.com/dcarne33/RNN-MC).

**Results**

*Recurrent and plain neural network comparison*

The plain NN and proposed RNN architecture and workflow are shown in Figs. 1(a-b). The RNN runs the four dimensionless optical properties of each layer and the hidden state through the network sequentially starting with the top layer. After each layer is run, the spectral response including the reflectance ($R$), absorptance ($A$), and transmittance ($T$) for the entire multi-layer medium is output. The hidden state ($h_i$) is initiated as all zeros for the first iteration of the RNN. Subsequent iterations use the hidden state output from the previous iteration of the RNN. The values of the hidden state passed between iterations are trained through the backpropagation process, meaning they are not manually selected. We theorize that the RNN is able to compress all necessary information about previous layers into a hidden state, allowing an RNN to be used for any number of layers instead of a separate plain NN for each number of layers. Training and validation datasets are generated using Monte Carlo simulations, as shown in Fig. 1(c), modeling 50,000 particles per simulation sampled over a broad range of optical properties. Using an RNN provides three key benefits over a plain NN which we demonstrate here.

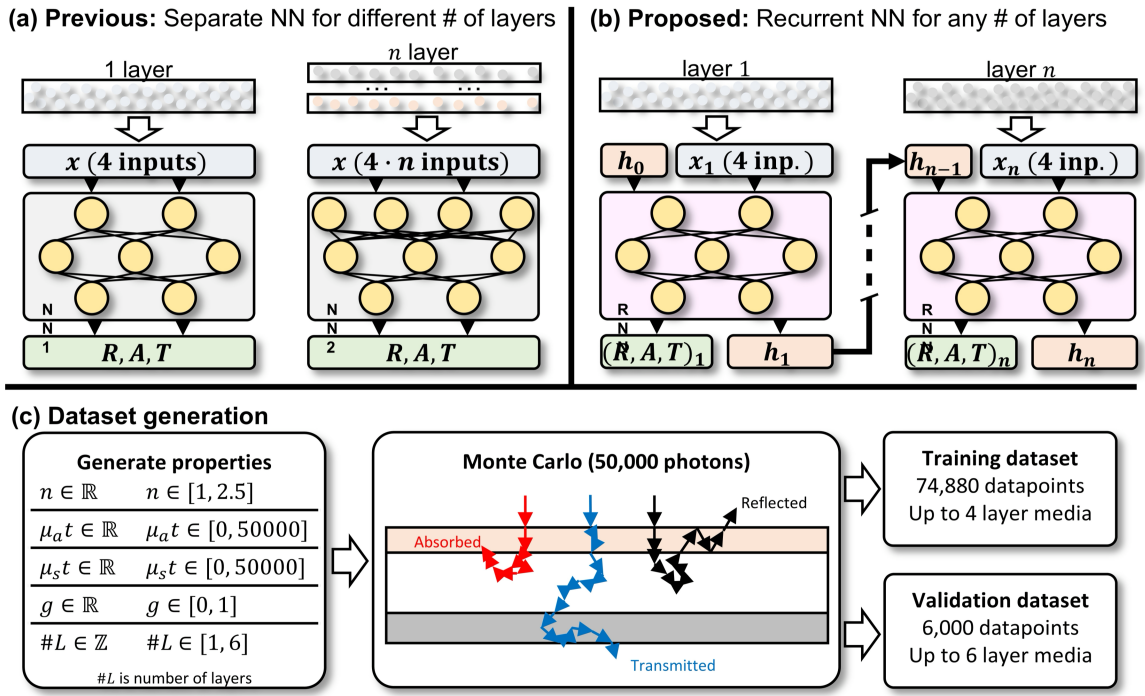

Figure 1: (a) Current process used for machine learning radiative predictions, where a separate NN is used for media with different numbers of layers. (b) Proposed solution using an RNN to recursively input layers, where $x_i$ is the input tensor, $h_i$ is the hidden state, and $(R, A, T)_n$ is the spectral response after $n$ layers. (c) Dataset generation using Monte Carlo simulations.

The first key benefit is that the RNN solves the curse of dimensionality for multi-layer radiative transfer problems. Using a plain NN requires four inputs for one layer, eight inputs for two layers, twelve inputs for three layers, and so on. As the number of input dimensions increases, the neural network requires exponentially more data to achieve comparable accuracy. This is a problem because the number of Monte Carlo simulations required for highly accurate multi-layer predictions becomes unfeasible. The RNN solves this problem as the number of inputs into the network, excluding the hidden state, remains at four regardless of the number of layers. This also makes intuitive sense that the RNN would perform better than a plain NN here. For a plain NN to learn four layer media, it has four separate input nodes for the refractive index of each layer. The plain NN needs enough data to learn how these four dimensions interact with all the other dimensions. As for the RNN, the refractive index is always input into the same node, so the RNN only needs enough data to learn the one dimension instead of four. To test our prediction that the RNN will significantly outperform the plain NN for the same dataset, four separate plain NNs are trained on the exact same dataset the RNN is trained on. Each plain NN has the same number of nodes and hidden layers as the RNN. Figure 2(a) shows the predicted vs true spectral response of the RNN and plain NNs, and Fig. 2(b) shows the mean prediction time as a function of mean absolute error for Monte Carlo, the RNN, and each plain NN. The RNN achieves between a significant 340-fold mean speedup over the Monte Carlo simulation (50,000 photons), at the cost of 3.9-fold increased error. Further comparison at equivalent levels of error is performed in the included case studies. Due to the curse of dimensionality, in Fig. 2(b) the plain NN faces an 8.1-fold increase in error in four layer media predictions compared to one layer media, whereas the RNN sees only a 1.5-fold increase. The same trend is observed in the maximum absolute error, where the RNN achieves 11.5-fold lower maximum error than the plain NN (0.040 vs 0.46) for four layer media. This data clearly shows the curse of dimensionality with the plain

NN suffering significant increases in mean and maximum error as the number of layers increases, whereas the RNN remains stable or marginally increases.

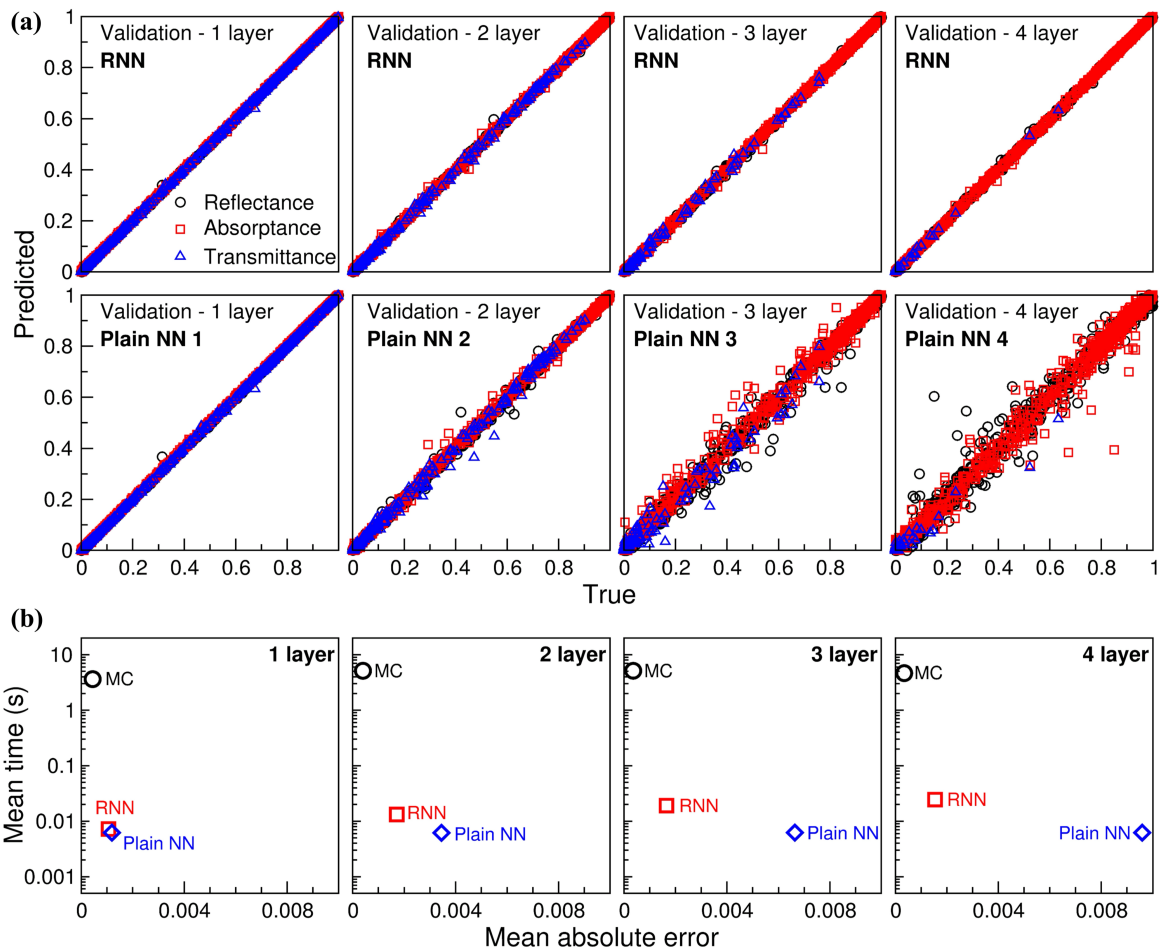

Figure 2: Comparison of RNN vs. plain NN on media composed of one to four layers on the validation dataset. (a) Predicted vs. true reflectance, absorptance, and transmittance. (b) Mean prediction time per datapoint as a function of mean absolute error.

The second key benefit is the ability of the RNN to generalize the multi-layer radiative transfer problem to predict any number of layers, even if the number of layers is outside the range of the training dataset. Figure 3(a) shows the predicted vs true spectral response of the RNN for five and six layer media, and the mean absolute error in Fig. 3(b). These results show that while the RNN was solely trained on data including up to four layer media, it successfully predicts five and six layer media with only marginally higher mean absolute error. Additionally, as highlighted in Fig. 3(b), enforcing physical laws, including the summation law ($R + A + T = 1$), after the prediction is made reduces the mean absolute error by 2.6%. The ability of the RNN to generalize the problem for any number of layers physically makes sense. Imagine a two-layer medium, where one knows the individual spectral response of each layer. These values cannot easily be combined, because the angular distribution of radiation between the two layers will change the spectral response of each layer, as well as the differences in Fresnel reflectance at the boundary. To combine the two layers into one medium, the radiation angular distribution at the bottom surface of the upper layer must be known. Once these two layers are combined, they can be considered as one effective layer. To add on a third layer, all that is needed is the spectral response of the effective layer (first two layers) and the radiation angular distribution of the bottom surface of the effective layer. The radiation angular distribution of each individual layer within the effective layer is not needed as that is accounted for in the effective layer's specular response. So, in theory, the hidden state would carry information including the spectral response of an effective layer including all the previous layers iterated through the RNN, and the radiation angular distribution of the bottom most surface of the effective layer. Since the spectral response of each layer can be compressed into a single effective layer, this prevents any long-term memory issues that may come up such as in natural language processing where more complex RNNs such as Long Short Term Memory (LSTM) RNNS and Gated Recurrent Units (GRUs) are used [29].

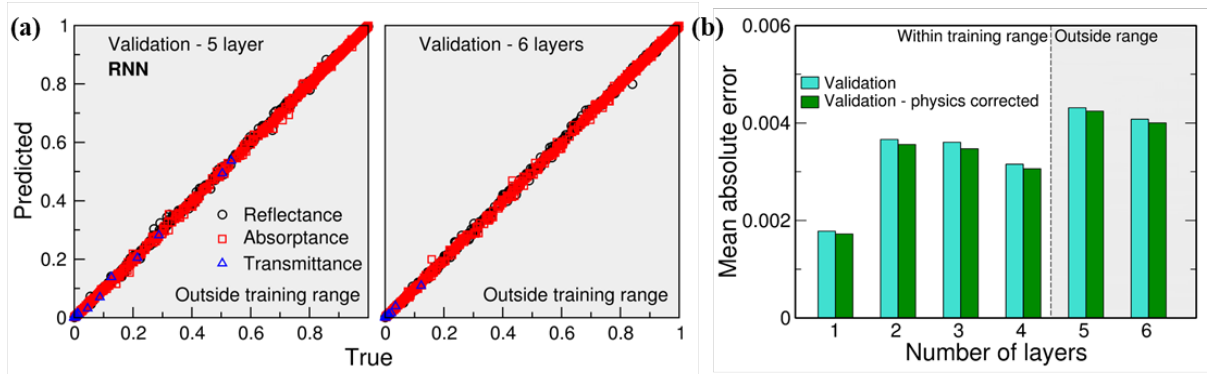

Figure 3: RNN accuracy on the validation dataset with up to six layer media. (a) Predicted vs. true reflectance, absorptance, and transmittance. (b) Mean absolute error for varying numbers of layers.

In Fig. 3(a), the datasets with five and six layer media has a mean transmittance of less than 1%. The argument could be made that the RNN only successfully predicts five and six layer media because those additional layers don't affect the spectral response. An additional validation dataset with 500 Monte Carlo simulations per number of layers for six through ten layer media is generated by sampling thinner layers to allow for high transmittance (mean transmittance greater than 39%). On this dataset (Figure S1(b)), while the error does increase gradually as the number of layers increase, even for ten layer media the average absolute error is less than 0.03 (3 percentage points), showing the RNN's ability to generalize the problem for media with any number of layers.

The third key benefit of the RNN architecture is the ability to train on the entire dataset. Even though the RNN and plain NN for one layer media have the same number of inputs and train on the same number of single layer datapoints, the RNN has 11% lower mean absolute error (Fig. 2(b)). This is because the RNN trains on the entire dataset (74,880 simulations) including one through four layer media, while each plain NN can only train on a set number of layers (18,720 simulations per number of layers). This reduction in error shows training the RNN on multi-layer media will also reduce the single layer error and vice-versa. It allows for considerably more efficient usage of the dataset compared to using a plain NN where the training dataset must be split up based on the number of layers for each network to train on.

To highlight the broad applicability of this RNN framework and to directly compare the RNN and plain NN to Monte Carlo simulations at a comparable accuracy, we present three case studies on multi-layer tissue, radiative cooling paint, and atmospheric models.

*Case study 1: Melanosome absorption in phototherapy*

Radiation plays many important roles in the medical field, largely through radiation therapy [30] and imaging techniques [31]. Within radiation therapy, Monte Carlo simulations are commonly used for dosimetry, to predict and optimize the absorption of radiation within tissue [32]. One such type of radiation therapy is phototherapy, which utilizes a 430 – 490 nm laser to treat a variety of conditions affecting the skin including hyperbilirubinemia [33]. Here, case study 1 (Figs. 4(a-b)) investigates absorptance of 480 nm light with varying melanosome volume fraction in skin tissue, considering the epidermis, dermis, and subcutaneous fat as three layers with differing optical properties. Details on optical property calculation and modeling can be found in section 5 of the SI. The RNN, plain NN, and Monte Carlo absorptance prediction as a function of melanosome volume fraction is seen in Fig. 4(a). Here, the RNN on average achieves 6.8 times lower error than the plain NN. Figure 4(b) plots the absolute error as a function of time per simulation on a log-log scale. Monte Carlo simulations can readily trade error for computational time by modeling an increased number of particles. Therefore, comparisons of the RNN to the Monte Carlo method must compare time savings at an equivalent error, and decreased error at an equivalent simulation time based on the trendline shown. Monte Carlo simulation with 2,000,000 particles simulated (mean error less than 0.0002) is considered the ground truth value for determining absolute error. For the same time required, the RNN provides 6.4x reduced error compared to Monte

Carlo. For the same error, the RNN provides a 43x acceleration compared to Monte Carlo. Additionally, for case study 1, the RNN error is less than that of Monte Carlo with 50,000 particles simulated (Fig. 4(b)), the number of particles used to generate the training dataset. While the other case studies have higher error than Monte Carlo with 50,000 particles simulated (Figs. 4(d), 4(f)), this highlights the potential for machine learning to denoise the training dataset and achieve higher accuracy than the training data with proper scaling (scaling law shown in Fig. S2).

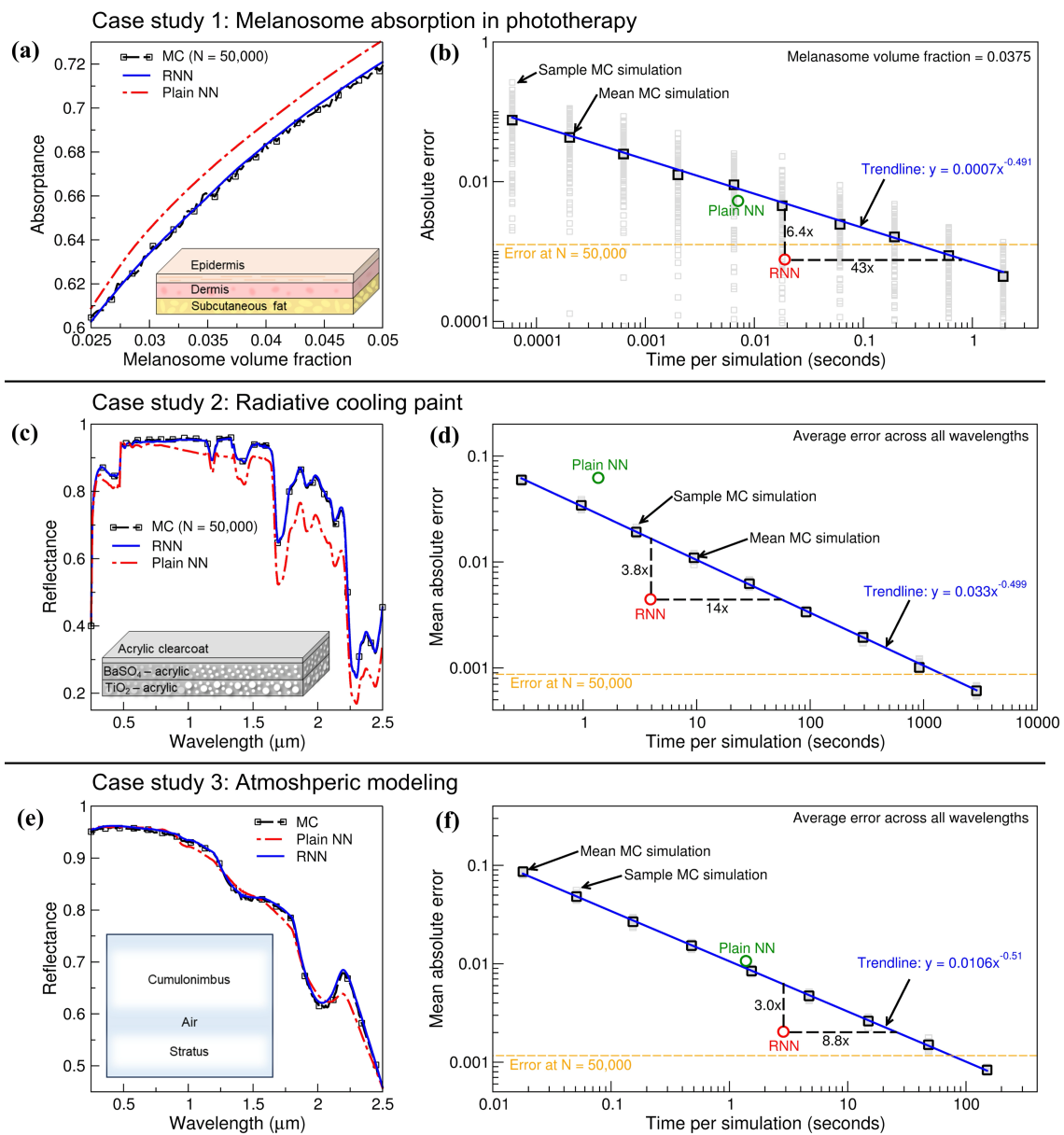

Figure 4: Case studies comparing the RNN to the Plain NN and Monte Carlo method for (a) three layer tissue, (b) three layer radiative cooling paint, and (c) two layer atmospheric clouds.

*Case study 2: Radiative cooling paint*

Spectrally selective radiative coatings are widely used across a broad range of applications including anti-reflection coatings for solar panels [34], highly reflective and variable emissivity coatings for spacecraft [35], highly absorptive coatings for solar heating [36], and low emissivity coatings for energy efficient windows [37]. Recently, radiative cooling paints have made significant advances, including the ability to provide full daytime sub-ambient cooling through a combination of high solar reflectivity and sky window emissivity [38], [39], [40]. Here, case study 2 (Figs. 4(c-d)) models the spectral reflectance of a three layer radiative cooling paint, including an acrylic clear coat for mechanical protection, a layer of $BaSO_4$-acrlic paint for UV reflectance, and a layer of $TiO_2$-acrylic paint for strong visible through near infrared reflectance. Details on optical property calculation and modeling can be found in section 6 of the SI. Here, the RNN on average achieves 14 times lower error than the plain NN. Figure 4(d) plots the absolute error as a function of time per simulation on a log-log scale. For the same time required, the RNN provides 3.8x reduction in error compared to Monte Carlo. For the same error, the RNN provides a 14x acceleration compared to Monte Carlo.

*Case study 3: Atmospheric modeling*

Simulating atmospheric radiative transfer is important for weather modeling, climate change predictions, and studying the potential of solar geoengineering. The atmosphere is comprised of many optically active features which induce scattering and absorption, including various cloud types, ozone ($O_3$), carbon dioxide ($CO_2$), and a mixture of aerosols. While these features have complex geometries, many climate models apply the plane-parallel assumption, also known as the independent column approximation [41], to simplify the analysis and reduce the computational load [42]. Here, case study 4 (Figs. 4(e-f)) models the spectral reflectance of a simplified atmospheric model considering two layers of clouds, an upper layer cumulonimbus cloud and a lower layer stratus cloud. Details on optical property calculation and modeling can be found in section 7 of the SI. Here, the RNN on average achieves 5.2 times lower error than the plain NN. Figure 4(f) plots the absolute error as a function of time per simulation on a log-log scale. For the same time required, the RNN provides 3.0x reduced error compared to Monte Carlo. For the same error, the RNN provides an 8.8x acceleration compared to Monte Carlo.

To allow readers to test and utilize this RNN, it has been open sourced in two methods. First, the feedforward function and the RNN's parameters including the weights and biases are available on GitHub (https://github.com/dcarne33/RNN-MC). The Python code here is easy to use and walks users through how to run the RNN. Second, the RNN has been implemented into FOS (https://github.com/FastOpticalSpectrum/FOS), our open-source code for modeling nanoparticle media, to allow users accelerate optimization and test multi-layer designs such as for radiative cooling paint. In addition to the three case studies highlighted, a comparison to an experimentally measured paint sample is included in section 8 of the SI to highlight the applicability of the RNN. The work has immediate benefits for simulating, optimizing, and high throughput screening of plane-parallel media, but we also hope it will inspire future work in the use of RNNs for three-dimensional media including for more complex atmospheric conditions, combustion, nuclear, and tissue modeling.

**Conclusion**

In summary, this study demonstrates a recurrent neural network to significantly accelerate single and multi-layer radiative transfer predictions typically calculated through Monte Carlo simulations. The RNN architecture provides three key benefits over a plain NN used in previous studies. First, by recursively inputting a layer's optical properties, the total number of input dimensions remains the same as the single layer case and doesn't increase with the number of layers as a plain NN would. This solves the curse of dimensionality, allowing the RNN to significantly outperform a plain NN on the same size dataset. Second, the architecture allows for any number of layers to be predicted, even if the number of layers is outside the maximum number of layers in the training dataset. Third, the RNN can train on the entire dataset of multi-layer media, whereas each plain NN can only train on a set number of layers. Three case studies are shown including multi-layer tissue, radiative cooling paint, and atmospheric clouds. These examples show the speedup ranges up to 1-2 orders of magnitude over Monte Carlo simulations depending on the scattering and absorption properties, with significantly lower error than the plain NN. The RNN has been made open source allowing others to utilize the pretrained model for greatly accelerating optimization and high throughput screening for applications including dosimetry, atmospheric studies, and spectrally selective radiative coatings.

**Methodology**

*RNN architecture and workflow*

Although RNNs are commonly used for natural language processing and time series data [29], spectral response predictions of multi-layer scattering media can also be thought of as a sequential problem. The overall workflow and architecture of the RNN used in this work is shown in Fig 1. First, the layer optical properties are non-dimensionalized to allow for a broader range of input values as well as reduce the number of inputs from five to four. The four inputs include the refractive index ($n$), the dimensionless absorption coefficient ($\mu_a * t$), the dimensionless scattering coefficient ($\mu_s * t$), and the asymmetry parameter ($g$). Each of these inputs are then normalized based on knowledge of how they scale and to keep each input value between zero and one (more on this in section 1 of the SI). Each layer's optical properties are sequentially input into the RNN starting with the top layer, along with the hidden state ($h_i$) which is initialized as all zeros. The RNN outputs the spectral response of the composite medium, based on the layer properties provided to the RNN and any preceding layers, and generates the hidden state to pass to the next layer's input. After all the layers have been sequentially run through the RNN, the spectral response is output, including the reflectance ($R$), absorptance ($A$), and transmittance ($T$). Finally, physical laws are manually enforced by correcting any values over one (100%) and any values below zero (0%), and by dividing by the summation of reflectance, absorptance, and transmittance to enforce the summation law ($R + A + T = 1$). Finally, physical laws are enforced through two mechanisms. First, any value over one (100%) is set to one, and any value below zero (0%) is set to zero. Second, the summation law ($R + A + T = 1$) is enforced by normalizing each value as shown in eq. 1 for the corrected reflectance ($R_{cor}$).

$$R_{cor} = \frac{R}{R + A + T} \qquad (1)$$

The RNN architecture includes three hidden layers with 1024 nodes each and 16 values in the hidden state. ReLU activation [43], [44] is applied to each hidden layer, the loss function is Mean Square Error (MSE) [44], and the optimizer is Adam [45] with a learning rate of 0.0001, with each parameter selected based on hyperparameter testing. The plain Neural Network (plain NN) is trained using the same number of nodes, hidden layers, optimization, and hyperparameters. Overall, so long as Adam optimization is used, any sufficient number of nodes, hidden layers, and hidden state sizes only marginally affects the validation error for both the RNN and Plain NN.

*Monte Carlo modeling and dataset generation*

Creating the dataset to train and evaluate the RNN and plain NN, as shown in Fig. 1(c), includes three main steps, selecting the range of optical properties, the sampling method, and the Monte Carlo modeling. A dataset of 80,880 Monte Carlo simulations with 50,000 particles per simulation is generated using FOS [46], our open source code where the Monte Carlo method is based on the open source code MCML by Wang et al. [12]. Monte Carlo modeling typically takes five inputs per layer, the refractive index ($n$), absorption coefficient ($\mu_a$), scattering coefficient ($\mu_s$), asymmetry parameter ($g$), and layer thickness ($t$). Here, to reduce the number of inputs into the neural network from five to four per layer, the dimensionless scattering coefficient ($\mu_s t$) and absorption coefficient ($\mu_a t$) are used and thickness removed from the inputs. These properties are randomly sampled with $n$, $g$, and the number of layers ($\#L$) being uniformly sampled, and $\mu_s t$ and $\mu_a t$ sampled on a logarithmic scale due to knowledge of how scattering and absorption coefficients scale (sampling detailed in section 2 of the SI). The data is then split so that there are 74,880 simulations in the training dataset containing multi-layer media uniformly sampled from one to four layers, and 6,000 simulations in the validation dataset containing multi-layer media uniformly sampled from one to six layers.

**Disclaimer**

# Supplemental Information for:

Overcoming the curse of dimensionality: Recurrent neural network for accelerated multi-layer radiative transport

Daniel Carne[a], Ziqi Guo[a], Xiulin Ruan[a,*]

[a] School of Mechanical Engineering, Purdue University, West Lafayette, IN 47907, USA

[*] Corresponding Author: ruan@purdue.edu

**1: Input normalization**

The input normalization formulas shown in Eqs. (1 – 4) are manually tested by trial and error and selected based off knowledge of how varying one input affects the spectral response. The neural network does not necessarily need to have the inputs normalized in this manner, but it does assist the network train faster as the solution is easier to fit. Additionally, each input is normalized between zero and one so that any one input does not have a higher initial weighting than others.

$$n_{norm} = n/2.5 \tag{1}$$

$$(\mu_a * t)_{norm} = (\log(\mu_a * t + 0.0001) + 4)/8.69154 \tag{2}$$

$$(\mu_s * t)_{norm} = (\log(\mu_s * t + 0.0001) + 4)/8.69549 \tag{3}$$

$$g_{norm} = 1 - (1 - g)^{1/3} \tag{4}$$

**2: Dataset sampling**

Data sampling for the refractive index and asymmetry parameter are linear. However, since scattering and absorption trend exponentially, the absorption coefficient, scattering coefficient, and thickness are sampled differently as shown in Eqs. (5 – 7) where $\zeta$ is a random number uniformly sampled from zero to one. For example, a non-dimensionalized absorption coefficient going from 0.1 to 0.2 can have a bigger impact on reflectance than going from 1000 to 2000. Because of this, many different orders of magnitude must be sampled.

$$\mu_a = 0.01(10^{8\zeta} - 1) \tag{5}$$

$$\mu_s = 0.01(10^{8\zeta} - 1) \tag{6}$$

$$t = 0.002 + 0.005(10^{1.0253\zeta} - 1) \tag{7}$$

## 3. RNN many layer predictions

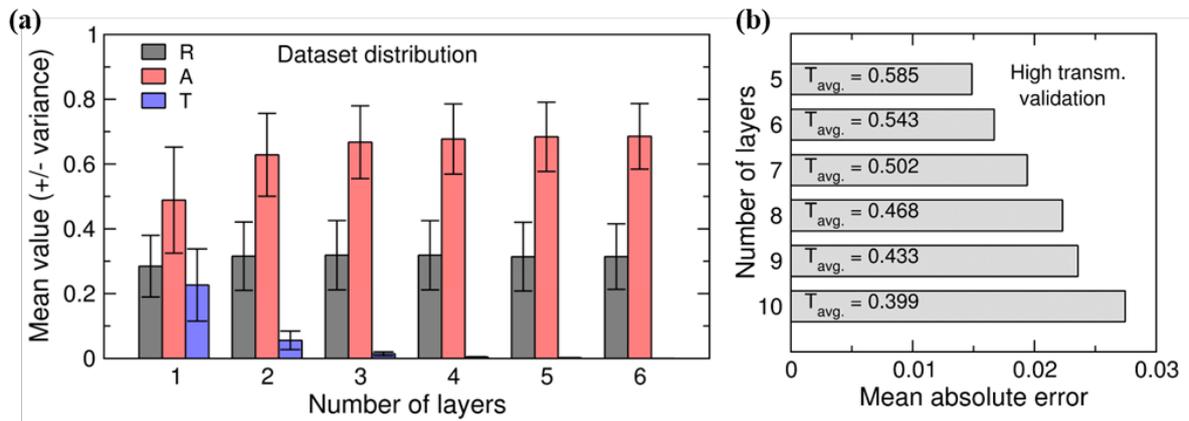

Figure S1: (a) Mean dataset spectral response (+/- variance) for varying numbers of layers. (b) RNN mean absolute error for five to ten layer media on a high transmission validation dataset.

## 4. Dataset scaling law for RNN

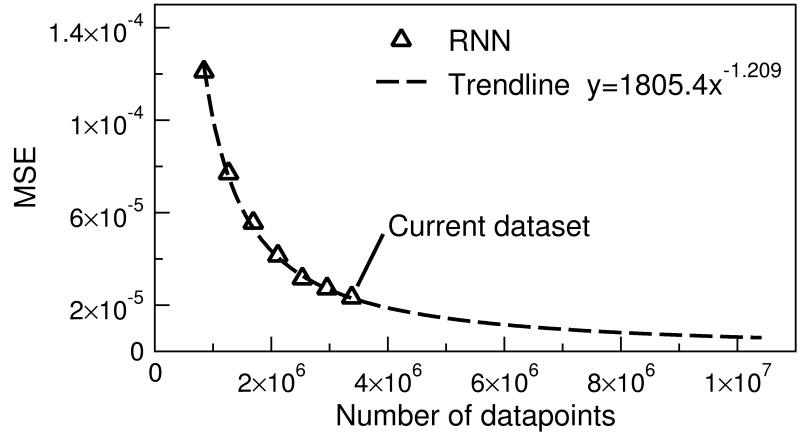

Figure S2: RNN mean square error as a function of dataset size with a power law trendline.

Figure S2 shows how the RNN Mean Square Error (MSE) scales with dataset size. This exhibits the RNN architecture follows a power law scaling relating MSE to dataset size, allowing for predictions of dataset size required to meet desired accuracy. This insight reveals that the RNN is not the limiting factor in accuracy but instead the dataset size and can be scaled based on user requirements.

## 5. Case study 1

The optical properties for each of the three layers are calculated from the equations and sources in Jacques et al. [1]. The reduced scattering coefficients for epidermis, dermis, and subcutaneous fat are from Salomatina et al. [2]. The refractive index is calculated based on the refractive index of water and dry tissue from Jacques et al. [3]. The percentage of blood, water, oxygen saturation, and fat to calculate the absorption coefficients for dermis and epidermis are from Choudhury et al. [4] and for subcutaneous fat from Weaver et al. [5]. The epidermis, dermis, and subcutaneous fat are from Lintzeri et al. [6], Ha et al. [7], and Kanehisa et al. [8], respectively. The optical properties chosen based on similar values to these sources are listed in Table S1. The epidermis absorption coefficient ($\mu_a = 500 f_{mel}$) is a function of the melanosome volume fraction ($f_{mel}$) which is varied from 0.025 – 0.05.

Table S1: Optical properties of three tissue layers including the refractive index ($n$), reduced scattering coefficient ($\mu_s'$), absorption coefficient ($\mu_a$), and layer thickness ($t$).

| Layer | $n$ [−] | $\mu_s'$ [cm$^{-1}$] | $\mu_a$ [cm$^{-1}$] | $t$ [cm] |
|---|---|---|---|---|
| Epidermis | 1.50 | 71.5 | $500 f_{mel}$ | 0.0075 |
| Dermis | 1.39 | 47.4 | 0.192 | 0.1 |
| Subcutaneous fat | 1.48 | 15.2 | 0.755 | 2 |

## 6. Case study 2

The three layer radiative cooling paint consists of a 30 μm acrylic clear coat for mechanical properties, a 100 μm layer of BaSO$_4$-acrlic paint for UV reflectance, and a 400 μm layer of TiO$_2$-acrylic paint for strong visible through near infrared reflectance. The BaSO$_4$ is at a 60% volume fraction and 0.2 μm in diameter. The TiO$_2$ is at a 60% volume fraction and 0.6 μm in diameter. The optical properties for each of these layers are calculated using the Mie theory implementation in FOS [9], our open-source code, from 0.25 μm to 2.5 μm wavelength.

## 7. Case study 3

The optical properties of the two cloud layers are based on the number of drops per unit volume, liquid water content, mode radius, and cloud height presented in Stephens et al. [10]. These values, as shown in Table S2, are then used to calculate the optical properties with the Mie theory implementation in FOS [9] from 0.25 μm to 2.5 μm wavelength.

Table S2: Properties of stratus and cumulonimbus clouds based on values from Stephens et al. [10].

| Layer | Number of drops [cm$^{-1}$] | Liquid water content [g cm$^{-1}$] | Mode radius [μm] | Thickness [km] |
|---|---|---|---|---|
| Stratus | 440 | 0.22 | 3.5 | 0.5 |
| Cumulonimbus | 72 | 2.50 | 6 | 3 |

## 8. Experimental comparison

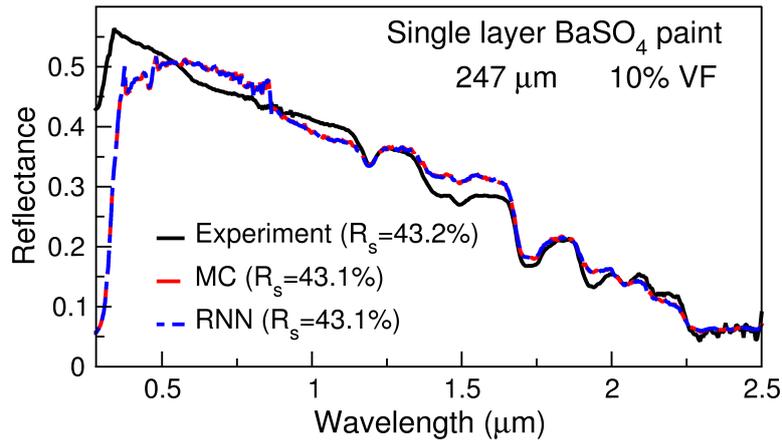

Figure S3: Reflectance as a function of wavelength for BaSO4 paint from experimental measurement, Monte Carlo simulation, and RNN prediction.

While the Monte Carlo method is a well validated method to solve radiative transport [11], we include a comparison to experimental results here to highlight the applicability of the RNN. A 247 μm thick paint with a silicone-based binder is manufactured at a 10% volume fraction of 398 ± 130 nm BaSO4 pigment. UV-Vis-NIR spectroscopy measured the spectral response from 0.25 μm to 2.5 μm wavelength, providing an integrated solar reflectance of 43.2%, whereas Monte Carlo and the RNN predict an integrated solar reflectance of 43.1%.